# Electrical spin injection into p-doped quantum dots through a tunnel barrier

L. Lombez [1], P. Renucci [1], P. Gallo [2], P.F. Braun [1], , H. Carrère [1], P.H. Binh [1,4], X. Marie [1], T. Amand [1], B. Urbaszek [1], J.L. Gauffier [1], T. Camps [2], A. Arnoult [2], C. Fontaine [2], C. Deranlot [3], R. Mattana [3], H. Jaffrès [3] and J.-M. George [3]

[1] *Laboratoire de Nanophysique Magnétisme et Optoélectronique – INSA, 135 avenue de Rangueil, 31077 Toulouse cedex 4, France*
[2] *Laboratoire d'Analyse et d'Architecture des Systèmes – CNRS, Université de Toulouse, 7 avenue du Colonel Roche, 31077 Toulouse cedex 4, France*
[3] *Unité Mixte de Physique CNRS/Thales, route départementale 128, 91767 Palaiseau cedex, France and Université Paris-Sud 11, 91405 Orsay, France*
[4] *Institute of Material Science, 18 Hoang Quoc Viet road, Cau Giay dist, Hanoï, Vietnam*

We have demonstrated by electroluminescence the injection of spin polarized electrons through Co/Al$_2$O$_3$/GaAs tunnel barrier into p-doped InAs/GaAs quantum dots embedded in a PIN GaAs light emitting diode. The spin relaxation processes in the p-doped quantum dots are characterized independently by optical measurements (time and polarization resolved photoluminescence). The measured electroluminescence circular polarization is about 15 % at low temperature in a 2T magnetic field, proving an efficient electrical spin injection yield in the quantum dots. Moreover, this electroluminescence circular polarization is stable up to 70 K.

The achievement of spintronic devices necessitates on the one hand the ability to inject spin polarized currents into semiconductors[1], and on the other hand sufficiently long spin relaxation times to manipulate or store the spin orientation of the injected carriers. The fundamental problem of conductivity mismatch between a ferromagnetic metal and a semiconductor pointed by *Schmidt et al.*[2] can be solved by the insertion of a tunnel barrier[3,4] which may be achieved either by a thin Schottky barrier[5,6], or by an insulating layer as aluminium oxide[6-8] or Manganese Oxide[9]. In addition to efficient injection, long electron spin relaxation time is a requirement for efficient storage and manipulation. This is indeed the case for semiconductor quantum dots[10]. Electrical spin injection into InAs/GaAs intrinsic QDs has been reported with spin aligner systems based on magnetic semiconductor[11-13] or ferromagnetic metal/semiconductor Schottky barrier.[14-15] In this letter, we demonstrate the efficient electrical spin injection of polarized electrons into p-doped InAs/GaAs quantum dots through an insulating tunnel barrier of Al$_2$O$_3$. These QDs are very promising for several reasons: (i) the polarization of p-doped QD luminescence is directly related to electron spin polarization even for zero applied magnetic field, because of the cancellation of the anisotropic exchange interaction in the positively charged exciton[16]; ii) very long electron spin relaxation times were reported recently[16] in these QDs under weak external magnetic field (about 100mT), which screens the effects of hyperfine interaction between electron and nuclear spins; (iii) long polarization decays (under excitation in the GaAs barrier) were measured even at high temperature.[17,18]

The Spin-LED (SLED) structure (Fig. 1) was grown by molecular beam epitaxy for the semiconductor part and by sputtering for the tunnel barrier/ferromagnetic spin injector part. It consists in a P-I-N GaAs structure grown on a p-doped GaAs:Zn substrate (p = 2. 10$^{18}$ cm$^{-3}$). The p-doped GaAs:Be layer (p = 1. 10$^{18}$ cm$^{-3}$) is 2 µm thick. The doping of the n-GaAs layer (thickness: 50 nm) is 10$^{16}$ cm$^{-3}$ to minimize the spin relaxation processes[19] at low temperature. The intrinsic GaAs zone contains five QD planes, separated by 30 nm in order to keep them electronically and mechanically uncoupled. A growth temperature of 500°C with a rate of 0.05ml/s was used resulting in a QDs density of the order of $10^{10}$ cm$^{-2}$. The thickness of deposited InAs corresponds to 10% more than the 2D-3D transition (about 1.8 monolayer). QDs have been doped with one hole per dot in average thanks to Be δ-doped planes located 15nm from the QD planes. The structure was capped by amorphous As. It was desorbed at 450°C in a chamber under ultra high vacuum in which desorption is monitored by RHEED. The sample was then transferred in situ in the sputtering chamber where $Al_2O_3$ tunnel barrier and cobalt ferromagnetic thin film (8 nm) were grown at room temperature. Tunnel barrier was formed by oxidation of 1.5 nm Al under a $O_2$ and Ar plasma. Gold cap layer (2 nm) was finally deposited to prevent cobalt from oxidation. The size of the contact area on the sample is 3*3 mm$^2$. We performed measurements on two samples S1 and S2. S1 is the structure described above, whereas S2 is a test structure without $Al_2O_3$ and Co for time and polarization-resolved photoluminescence (TRPL) experiments. For the electroluminescence (EL) measurement, the SLED was placed into a magnetic field (superconducting coil) in a cryostat with an exchange gas temperature ranging from 1.7 K to 70 K. The device was forward biased with square voltage pulses of 5 μs width, and a repetition rate of 50 kHz. The EL signal was detected in the Faraday geometry by a standard photoluminescence (PL) set-up (applied magnetic field along the growth axis). For TRPL experiments, a mode-locked Ti:sapphire laser (1.5 ps pulse width) was used for the non-resonant circularly-polarized excitation at ~1.549 eV (i.e. in the GaAs barrier) at 10 K. The PL signal was detected by a synchroscan streak camera, which provides an overall temporal resolution of 30 ps. The circular polarization was analyzed by passing EL and PL through a λ/4 wave plate and a linear analyzer.

The quantum dots embedded in the PIN junction were first characterized by TRPL. The photoluminescence circular polarization rate $P_C = (\Gamma^+ - \Gamma^-)/(\Gamma^+ + \Gamma^-)$ is time-resolved ($\Gamma^+$ and $\Gamma^-$ are the intensities of the right and left circularly polarized components of the luminescence respectively). We have checked (not shown) that, after an excitation in the wetting layer (~1.441 eV), $P_c$ exhibits the same behaviour as in ref [16], where a polarization around 15% is observed at long time delay without any magnetic field. It indicates that our sample is very comparable to the ones of ref [16]. In these structures, a very long electron spin relaxation time $\tau_s$ larger than 4 ns under a small longitudinal magnetic field B superior to 100 mT was measured. This slow relaxation is due to the quenching by the applied magnetic field of the effect of the hyperfine interaction between electron and nuclear spins, responsible for the electron spin relaxation in these systems. Fig. 2 displays the kinetics of the two polarized components of luminescence and the corresponding circular polarization rate after a non resonant laser excitation in the GaAs barrier (at ~1.549 eV) at B = 0. These conditions of optical excitation are close to the one encountered for an electrical injection of carriers, where the electrons are also non resonantly injected in the QDs from the barrier. A slow decay of the polarization rate is measured, with a characteristic decay time of ~ 2 ns. Note that, as said previously, this decay should be even slower when a small longitudinal magnetic field superior to 100 mT is applied, which is the case in the standard conditions of operation of the device. As a consequence, the polarization decay is much slower than the carrier lifetime, and one can thus consider[20] that the electron spin relaxation is negligible once the electron is trapped on the fundamental level ($S^e$) in the dot. It makes these dots very suitable for probing the electron spin polarization immediately after this trapping.

We measure (see Fig.3a and Fig3c.) the electroluminescence circular polarization ($P_C^{EL}$) at the peak energy corresponding to QDs emission of about 15% for B = 2T. Note that we have checked that the magnetic circular dichroism is negligible (inferior to 2%) by a photoluminescence experiment under linearly polarized non resonant excitation in the GaAs barriers (not shown). The spurious effect of Zeeman splitting in the dots can also be estimated to be inferior to 1% for this range of magnetic field and temperature[13]. In addition to QD emission, a spectral feature is observed at 1,49 eV in Fig 3a., attributed to a conduction band-neutral acceptor transition (e-$A_0$) probably due to Beryllium impurity [21] in GaAs. This optical transition constitutes an interesting additional probe of the electron spin polarization, directly in the GaAs after injection through the tunnel barrier. Due to degeneracy[22], we cannot distinguish between the heavy and light holes states for the impurety, and the electron spin polarization in GaAs should be twice[23] the circular polarization rate of about 13 % measured for the luminescence of the transition (e-$A_0$) for B = 2T. In the same time, the peak energy corresponding to QDs emission ($S^e$-$S^{hh}$ transition) exhibits an EL circular polarization of about 15%. For QDs, the polarization rate of luminescence should directly reflect the electron spin polarization in the dots, thanks to the lift of degeneracy between heavy and light holes in these systems. The electrically injected electrons spin polarization seems thus lower in the dots (about 15 %) than in GaAs (about 26%). A spin relaxation during the trapping into the dots may explain this discrepancy, but another argument should also be taken into account : under CW electrical excitation in the range of current density employed here (4.3 to 7.5 A.cm$^{-2}$), part of the electroluminescence may arise not only from positively charged, but also from neutral, negatively and multi-charged dots [24,25]. These additional contributions lead mainly to a reduction of the averaged EL circular polarization rate measured on an ensemble of quantum dots[18]. Fig.3c displays the magnetic field dependence of the EL polarization rate for QDs and (e-$A_0$) transitions. It follows for both transitions the Co layer saturation along the hard axis of magnetization of the layer. For B=2T, the electron spin polarization $P_{Co}$ in the Co layer is about 42%[26]. The ratio $P_C^{EL}/P_{Co}$, which constitutes a simple figure of merit to compare different systems, is at least 35% inside the dots. In this analysis, we have neglected spin relaxation in n-doped GaAs layer, which is minimized at low temperature by the appropriate n-doping level of n ~ $10^{16}$ cm$^{-3}$ [19]. Fig 4 displays the EL circular polarization when the temperature is increased. We found that $P_C^{EL}$ is stable up to 70 K (Fig 4.a). The low signal level for higher temperature prevents us from probing spin injection out of this range. Finally, we observe that the EL circular polarization is robust in a range of electric current density from 4.3 to 7.5 A.cm$^{-2}$ (see Fig 4b.).

We have demonstrated by electroluminescence an efficient electrical spin injection through Co/$Al_2O_3$/GaAs structures into p-doped InAs/GaAs quantum dots embedded in a PIN GaAs light emitting diode. The spin polarized electrons are stored in quantum dots and the spin information can be read by radiative recombination. This device gets benefits from the very slow electron spin relaxation time within these quantum dots. The experimental circular polarization of electroluminescence is about 15 % up to 70 K under 2T. The electrical spin injection in the quantum dots is characterized by a figure of merit estimated to be at least 35%, considering an electron spin polarization of about 42% for Co.

This work is partially supported by the french Agence Nationale pour la Recherche (ANR) contract MOMES.

**Figure 1**

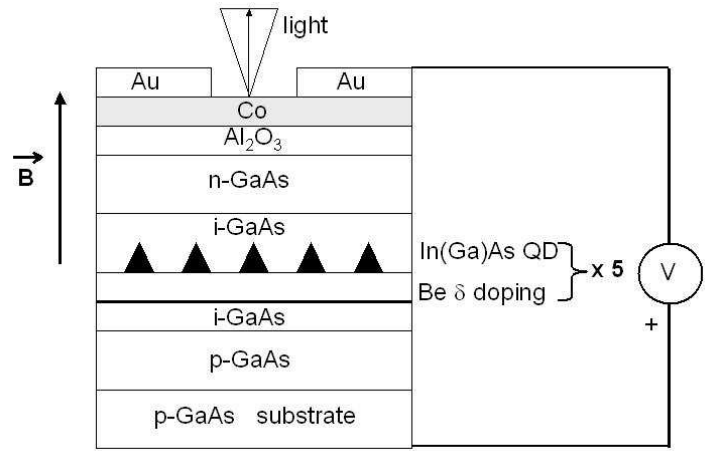

*Spin LED device with p-doped In(Ga)As quantum dots.*

**Figure 2**

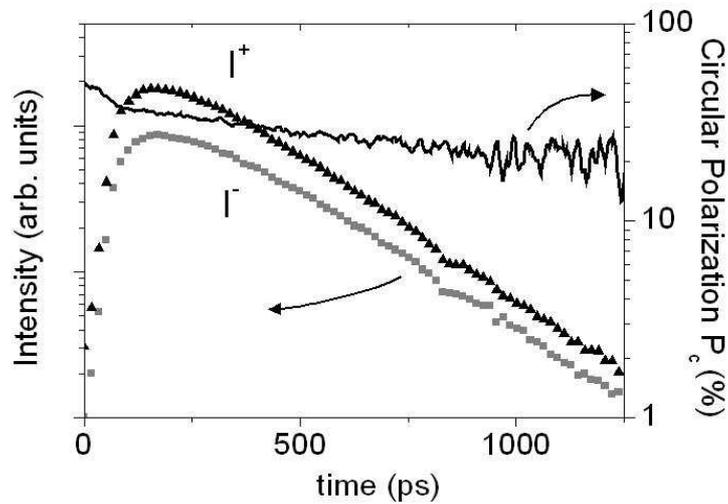

*Left axis: $I^+$ (black) and $I^-$ (grey) intensity components of PL (co- and counter-polarized with the laser) versus time at 10K for sample S2. Right axis: Corresponding time-resolved circular polarization rate; logarithmic scale.*

**Figure 3**

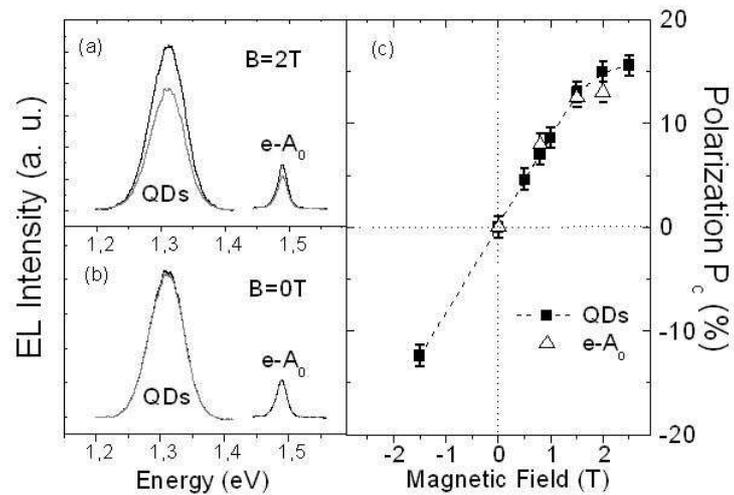

*a) For B = 2T EL spectra $I^+$ (black) and $I^-$ (grey) for the e-$A_0$ optical transition and for the QD luminescence peak (sample S1). The current density in SLED is 6. 3 A.cm$^{-2}$.*
*b) B= 0 T. $I^+$ and $I^-$ components of EL (sample S1).*
*c) Circular polarization rate of EL as a function of magnetic field at 1.7K (sample S1). Bold squares for QDs and empty triangles for the e-$A_0$ transition.*

**Figure 4**

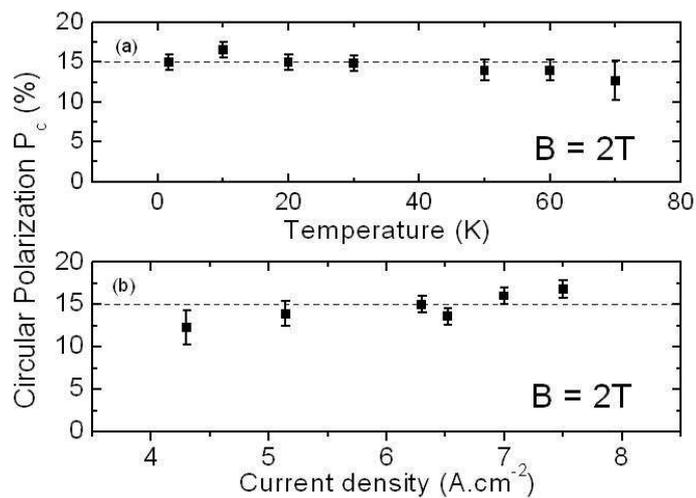

*(a) Circular polarization rate of EL as a function of temperature at B = 2 T(sample S1).*
*(b) Circular polarization rate of EL as a function of current density (T = 1,7K, B = 2 T)*